%% file: main.tex
\documentclass{article}

\usepackage{PRIMEarxiv}

\usepackage[utf8]{inputenc} 
\usepackage[T1]{fontenc}    
\usepackage{hyperref}       
\usepackage{url}            
\usepackage{booktabs}       
\usepackage{amsfonts}       
\usepackage{nicefrac}       
\usepackage{microtype}      
\usepackage{fancyhdr}       
\usepackage{graphicx}       
\usepackage{xcolor}
\usepackage{enumitem}
\usepackage{amsmath}
\newcommand{\comment}[1]{}
\usepackage{microtype}
\usepackage{booktabs}
\usepackage{tabularx}
\newcolumntype{Y}{>{\centering\arraybackslash}X}
\graphicspath{{media/}}     
\def\etal{\emph{et al}}
\definecolor{tabG}{HTML}{004d00}
\definecolor{tabR}{HTML}{de3238}
\title{Discriminative Adversarial Privacy: Balancing Accuracy and Membership Privacy in Neural Networks
}

\author{
  Eugenio Lomurno, Alberto Archetti, Francesca Ausonio, Matteo Matteucci \\
  Politecnico di Milano \\
  Milan, Italy\\
  \texttt{\{eugenio.lomurno, alberto.archetti, francesca.ausonio, matteo.matteucci\}@polimi.it} \\
}

\begin{document}
\maketitle


\input{src/abstract}


\input{src/introduction}

\input{src/related}

\input{src/method}

\input{src/experiments}

\input{src/results}

\input{src/conclusion}

\input{src/acknowledgements}


\bibliographystyle{unsrt}  
\bibliography{references}

\end{document}

%% file: src/abstract.tex
\begin{abstract}
The remarkable proliferation of deep learning across various industries has underscored the importance of data privacy and security in AI pipelines. As the evolution of sophisticated Membership Inference Attacks (MIAs) threatens the secrecy of individual-specific information used for training deep learning models, Differential Privacy (DP) raises as one of the most utilized techniques to protect models against malicious attacks. However, despite its proven theoretical properties, DP can significantly hamper model performance and increase training time, turning its use impractical in real-world scenarios. Tackling this issue, we present Discriminative Adversarial Privacy (DAP), a novel learning technique designed to address the limitations of DP by achieving a balance between model performance, speed, and privacy. DAP relies on adversarial training based on a novel loss function able to minimise the prediction error while maximising the MIA's error. In addition, we introduce a novel metric named Accuracy Over Privacy (AOP) to capture the performance-privacy trade-off. Finally, to validate our claims, we compare DAP with diverse DP scenarios, providing an analysis of the results from performance, time, and privacy preservation perspectives.
\end{abstract}

%% file: src/introduction.tex
\section{Introduction}\label{Introduction}
The burgeoning interest and application of deep learning across diverse industries and domains has been remarkable in recent years. This surge can be ascribed to several pivotal factors including the accessibility of massive data volumes, the enhancements in computational resources, and the evolution of neural network architectures and optimization algorithms. However, with the widespread use of deep learning models and their increasing influence on society and daily life, the security and protection of the sensitive data used to train such models have become an essential concern.
In an era of stringent data protection legislations like the European General Data Protection Regulation~\cite{albrecht_how_2016} and the Chinese Cyber Security Law~\cite{parasol_impact_2018}, threats and potential data security breaches evolve at a pace that is often faster than legislative response. One such threat is a class of attacks known as Membership Inference Attacks (MIAs)~\cite{shokri2017membership}, which aim to deduce whether certain individual-specific information was part of the training dataset of a machine learning model. Such attacks pose a formidable challenge and lay the ground for more sophisticated and potentially harmful breaches. Despite their recognition in the literature, no formally effective countermeasure is widely adopted in the deep learning development community.

Among the different solutions to increase the resistance of machine learning models to MIAs, the incorporation of Differential Privacy (DP) within training optimisers has stood out for its potential. DP is frequently considered the go-to mechanism for guaranteeing privacy due to its theoretical properties and robustness~\cite{abadi2016deep}. However, research has shown that the high level of privacy offered by DP comes at a cost, as adopting a high level of DP usually results in severe performance loss and increased learning time. This makes DP not practical for many real-world applications and sometimes not even for simple simulations~\cite{lomurno2023utility,bagdasaryan2019differential}.

In this paper, we introduce a novel privacy-preserving learning technique, called Discriminative Adversarial Privacy (DAP). DAP leverages the structure of MIAs to accomplish multi-objective adversarial learning. By using our approach, the training process of deep learning models is faster than training with DP and the final model has higher performance with a comparable privacy level. Specifically, DAP employs a discriminator trained via the MIA technique of shadow models and a novel loss function minimising the prediction error while maximising the attacker's error. This approach results in models that offer privacy competitive to those achieved by DP, yet with significantly reduced performance loss and faster training time.

With this work, we provide the following contributions:
\begin{itemize}[topsep=0pt,itemsep=0pt,parsep=0pt,partopsep=0pt]
\item We propose a novel learning technique called Discriminative Adversarial Privacy or DAP, that combines adversarial learning and membership inference attack principles. This technique is designed to ensures an optimal balance between model performance, speed, and privacy.
\item We introduce a novel loss function for DAP that is specifically tailored to simultaneously minimise the prediction error while maximising the attacker's error.
\item We define a novel metric, namely Accuracy Over Privacy or AOP, to efficiently capture and handle the performance-privacy trade-off.
\item We substantiate our claims with rigorous empirical validation, providing extensive experimental results that demonstrate DAP's comparative advantage over DP in terms of performance, training time, and privacy preservation.
\end{itemize}

%% file: src/related.tex
\section{Related Works}\label{Related}

\textbf{Membership Inference Attacks.} The family of attacks known as Membership Inference Attacks (MIAs) is one of the biggest threats to deep learning models. MIAs are incredibly versatile and effective, leading to a growing research interest both in terms of the development of new attack algorithms and defensive countermeasures~\cite{hu2022membership}. MIAs consist of determining, given a machine learning model, whether or not a given record was included in its training dataset.
In practice, a MIA model is a binary classifier that can distinguish whether or not a record belongs to the training set of an already trained target model. The challenge is to carry out the MIA in the real world with little useful information for the attacker, such as in machine-learning-as-a-service scenarios. 
Shokri \etal~\cite{shokri2017membership} pioneered one of the first and, still to this day, highly effective MIA algorithms, based on the assumption that over-parameterised models could memorise information about individual training samples beyond the generalisation of the problem for which they were trained. Assuming the structure and learning algorithm of the target model are known to the attacker, Shokri \etal propose a training technique that trains several models -- called \emph{shadow} models -- to emulate the target model's behaviour. In this way, the attacker can leverage the predictions of such models to build a MIA discriminator, able to identify whether a sample has been used or not in the training procedure of the target model.

Numerous studies extended this technique and expanded the attack surface. For instance, Chen \etal used data poisoning to enhance the MIA precision while hiding the attack traces by minimising test-time performance degradation~\cite{chen2022amplifying}. He \etal demonstrated the feasibility of MIA against models trained via self-supervised learning, and explored early stopping as a potential countermeasure, albeit at the expense of the model's utility~\cite{he2022semi}. Recently, researchers evaluated the effectiveness of MIAs against Generative Adversarial Networks (GANs)~\cite{webster2021person,hu2021membership}, diffusion models~\cite{duan2023diffusion,hu2023membership}, recommender systems~\cite{wang2022debiasing,yuan2023interaction}, semantic segmentation~\cite{chobola2022membership,zhang2022label}, and text-to-image~\cite{wu2022membership}.

\par\medskip
\noindent\textbf{Differential Privacy.} Historically, Differential Privacy (DP) has been the primary defence against MIAs. It is a procedure designed to provide robust protection for individual-level information in a dataset~\cite{dwork2008differential}. The application of DP ensures that the inclusion or exclusion of any individual sample in a dataset does not significantly alter the results of statistical analyses or machine learning models trained on that dataset.
DP is frequently presented in its relaxed form, referred to as ($\varepsilon,\delta$)-DP. Formally, a randomised mechanism denoted as \emph{M: $D \rightarrow R$}, with domain \emph{D} and range \emph{R}, satisfies ($\varepsilon,\delta$)-DP if the following inequality holds for any two adjacent inputs $\emph{d, d'} \in \emph{D}$ and any subset of outputs $\emph{S} \subseteq \emph{R}$:
\input{eqs/dp}%
In Equation~\eqref{eq:dp}, the $\varepsilon$ parameter, known as the privacy budget, denotes the maximum allowable information leakage, with a lower $\varepsilon$ value indicating stronger privacy. Conversely, the additive $\delta$ term represents the probability of privacy preservation being violated.

In a machine learning context, the DP framework achieves its goal by introducing randomness into the data analysis process, in a manner that obscures the contribution of any single individual's data. Usually, this randomisation is implemented as additive noise summed to the original data, to the intermediate matrices of the training algorithm, or through ad-hoc subsampling of the dataset.
Abadi \etal~\cite{abadi2016deep} pioneered the concept of Differentially-Private Stochastic Gradient Descent (DP-SGD), which has been affirmed as one of the most prevalent differentially-private optimisers within the deep learning literature. DP-SGD introduces Gaussian noise to the gradient computation with a standard deviation controlled by $\varepsilon$. This step ensures that the gradients are sufficiently randomised, thereby hindering an attacker's ability to infer information about individual data points from the model's parameters.
This approach has been successfully applied across various domains, particularly in federated learning applications~\cite{lomurno2021sgde,liu2022privacy,adnan2022federated}.

\par\medskip
\noindent\textbf{Alternative Privacy Preserving Techniques.}
While DP offers numerous advantages such as increased trust in data analysis results and enhanced fairness in decision-making processes that rely on data, it does come with its own set of challenges. These primarily revolve around the trade-off between privacy protection and data utility and the computational complexity that arises while implementing DP mechanisms~\cite{lomurno2023utility,bagdasaryan2019differential}. Given these constraints, the literature has seen the emergence of alternatives to DP and its variants.
Chen \etal proposed an alternative to DP, called RelaxLoss~\cite{chen2022relaxloss}. The key concept of this training framework is the relaxation of the entropy loss function, with the goal of reducing the generalisation gap and privacy leakage in machine learning models.
Kaya and Dumitras~\cite{kaya2021does} evaluated the efficacy of data augmentation mechanisms against MIAs in image classification tasks. Their study encompassed seven different mechanisms, including differential privacy. They found that augmenting data to improve model utility did not mitigate the risk of MIAs. Furthermore, they delved into why the commonly utilised label smoothing mechanism amplified the risk of MIAs.
Webster \etal introduced a general-purpose approach to tackle the issue of membership privacy in machine learning. Their solution involves the generation of surrogate datasets using images created by Generative Adversarial Networks (GANs), which are labelled with a classifier trained on the private dataset. They demonstrated that these surrogate datasets can be utilised for various downstream tasks and provide resistance against membership attacks. In their study, different GANs proposed in the literature were evaluated, revealing that GANs of higher quality yield better surrogate data for the given task~\cite{webster2021generating}.
Lomurno and Matteucci~\cite{lomurno2023utility} presented a comparison of the effectiveness of the DP-SGD algorithm against standard optimisation practices with regularisation techniques. They compared the utility of the resulting models, their training performance, and the efficacy of MIAs against the learned models. Their empirical findings highlight the often superior privacy-preserving properties of dropout and $l2$-regularisation, given a fixed number of training epochs.

%% file: eqs/dp.tex
\begin{equation} 
    Pr[\emph{ M(d)} \in \emph{S }] \leq \emph{e}^\varepsilon Pr[\emph{ M(d')} \in \emph{S }] + \delta.
\label{eq:dp}
\end{equation}

%% file: src/method.tex
\section{Method}\label{Method}

\begin{figure}[t]
    \centering
    \includegraphics[width=\textwidth]{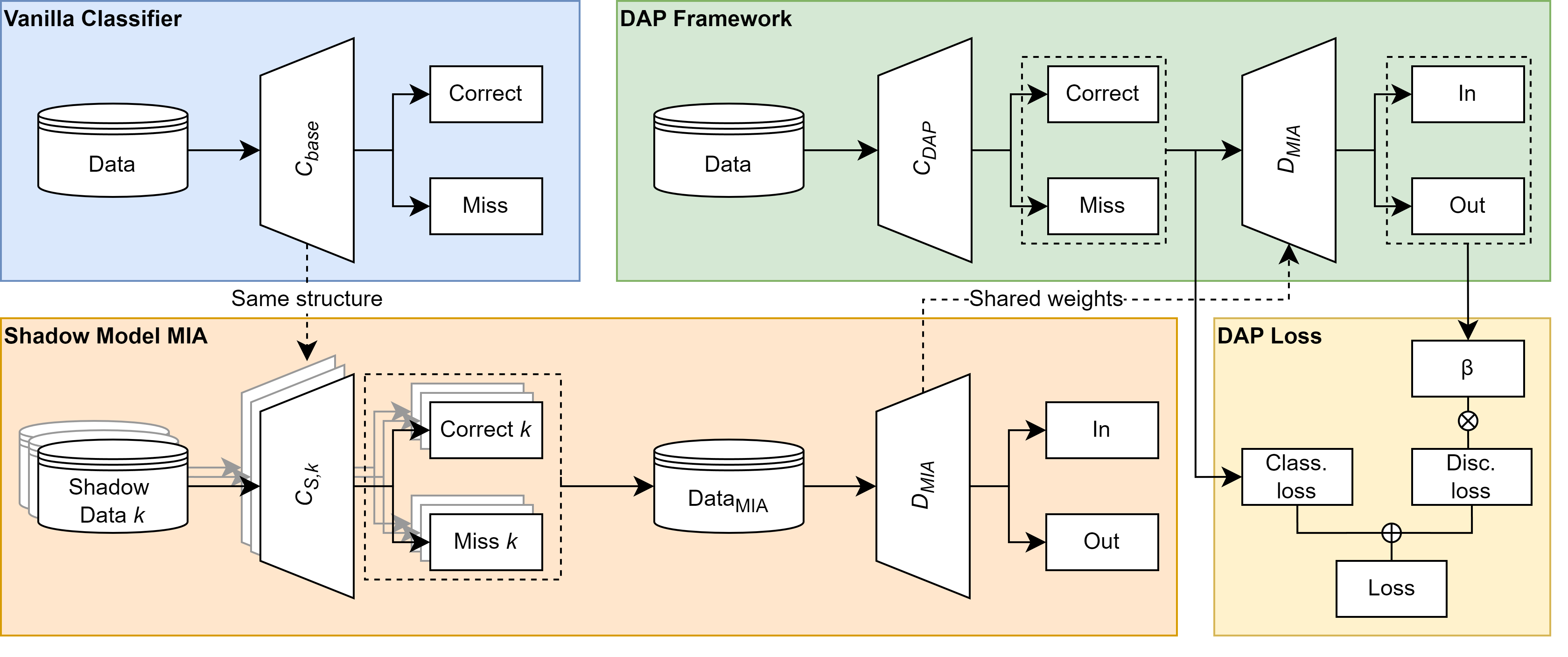}
    \caption{Overview of the Differentiable Adversarial Privacy (DAP) framework.}
    \label{fig:dap}
\end{figure}

\textbf{Discriminative Adversarial Privacy.}
With this work, we introduce a novel learning framework for privacy-preserving deep learning, called Discriminative Adversarial Privacy (DAP). 
DAP is a learning framework to efficiently train high-performing deep learning models with strong resilience against MIAs. 
DAP uses a deep neural network classifier as a baseline, referred to as $\mathcal{C}_{base}$, and trained using the hold-out technique on dataset \textit{Data}. 
Similarly, $K$ shadow models, denoted as $\mathcal{C}_{S,k}$, are trained using the hold-out technique, as prescribed by the MIA from Shokri \etal~\cite{shokri2017membership}.
For each shadow model produced this way, ground truth, prediction, and loss are stored for each of its training and test samples, associating a binary label according to whether it belongs to the first or second set.
Of these samples, only the miss-classified ones are retained, as they are the most empirically informative in a discriminative context and, from the ablation studies, lead to the most performing results.
These data are used to build the adversarial binary classification dataset \textit{Data}$_{MIA}$ to train the binary discriminator $\mathcal{D}_{MIA}$.
Once trained, the weights of this model are frozen.

At this point, adversarial training is performed using $\mathcal{D}_{MIA}$ and a new classifier $\mathcal{C}_{DAP}$ with the same structure of $\mathcal{C}_{base}$.
In DAP, $\mathcal{C}_{DAP}$ is trained to minimize the categorical crossentropy loss as usual, i.e. to maximize the probability of assigning the correct class label to training examples.
This error is used to update all the classifier's weights.
Then, for each batch of data, the miss-classified predictions from $\mathcal{C}_{DAP}$ are collected together with their corresponding ground truth and loss. 
This secondary batch is fed through $\mathcal{D}_{MIA}$ and its prediction error is computed maximising the error of the discriminator as in the standard min-max adversarial training~\cite{goodfellow2020generative}.
This secondary error is used to update the last fully connected layer of $\mathcal{C}_{DAP}$, with the goal of reducing the probabilities that its outputs can be easily discriminated by the attacker.
The optimisation procedure of DAP can be described as
\input{eqs/loss}%
In Equation~\eqref{eq:loss}, $x$ and $y$ are respectively the training inputs and the ground truth labels, $t$ is the current epoch, and $\beta$ is a dynamic loss balancing parameter.
$\beta$ is crucial to ensure learning stability. In fact, the different nature of the two losses makes them not directly comparable in terms of magnitude depending on the training epoch and the specific data distribution. $\beta$ is dynamically adjusted during training and it is computed as

\input{eqs/beta}According to Equation~\eqref{eq:beta}, the value of $\beta$ at time $t$ is proportional to the ratio of the classification loss and the discrimination loss on the validation set at the previous step $t-1$. Then, $\beta$ is scaled by a hyperparameter $r$ that weighs the contribution of the discriminator. As a final note, $\beta$ is always set to $1$ for $t = 0$. The overall DAP framework is described in Figure~\ref{fig:dap}.

\par\medskip
\noindent\textbf{Accuracy Over Privacy.}
When evaluating machine learning models in a privacy-preserving setting, it is vital to contemplate both model performance and the privacy of the underlying training data. However, measuring the trade-off between these two facets is a complex task. Quantifying privacy itself is nontrivial, and comparing metrics across different domains can be particularly challenging. For these reasons, we propose a novel metric called Accuracy Over Privacy (AOP), which provides a concise measure of the accuracy and privacy of a target model. 
Within the realm of MIAs, the efficacy of the attacking model is often measured using the Area Under the Curve (AUC) of the Receiver Operating Characteristic (ROC) curve of the attack. Instead, when the model is a classifier, its performance can be assessed utilising the Top-1 Accuracy (ACC). Therefore, given the classifier ACC and the AUC of a MIA model (AUC$_{\text{MIA}}$), the AOP is computed as

\input{eqs/aop}%
Equation~\ref{eq:aop} summarizes the effects of ACC and AUC$_\text{MIA}$ in a single metric. $\lambda \geq 1$ weighs the importance of privacy when measuring the AOP.

\begin{figure}[t]
    \centering
    \includegraphics[width=0.245\textwidth]{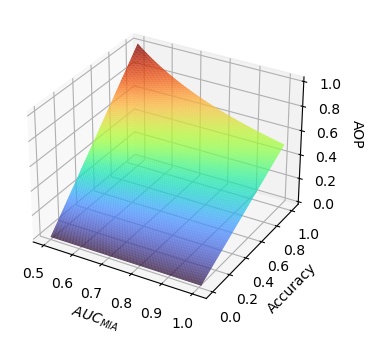}
    \includegraphics[width=0.245\textwidth]{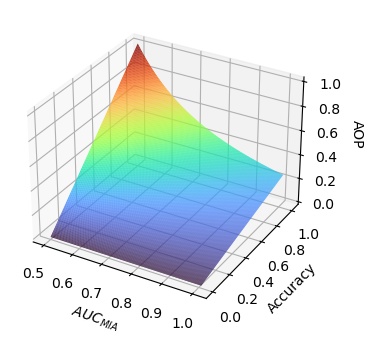}
    \includegraphics[width=0.245\textwidth]{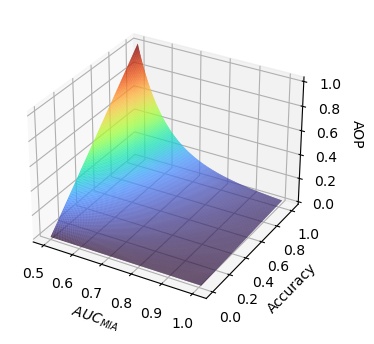}
    \includegraphics[width=0.245\textwidth]{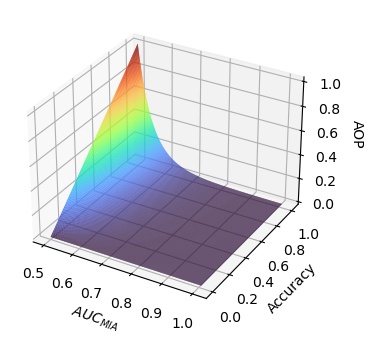}
    \caption{From left to right, interpolation plots of AOP($\lambda$) for $\lambda = 1$, $2$, $5$, and $10$.}
    \label{fig:aop}
    \vspace{-10pt}
\end{figure}

The AOP metric exhibits several properties. Concerning its range, it is constrained in the interval $[0,1]$. For highly inaccurate models or models susceptible to MIAs, the AOP approaches 0. Conversely, the AOP is closer to 1 when models exhibit high accuracy and strong resilience against MIAs at the same time.
The $\lambda$ parameter is a key factor in the AOP metric, as it is controls the impact of the privacy component. Figure~\ref{fig:aop} shows that increasing values of $\lambda$ cause the AOP metric to shrink towards 0.
Concerning the denominator, the $\text{max}$ operator ensures that the AUC is never lower than the AUC of a random guessing model, which is equal to 0.5. Moreover, the denominator allows for obtaining AOP values equal to the classification accuracy for models that perfectly preserve privacy.

%% file: eqs/loss.tex
\begin{equation}
\displaystyle\min_{\mathcal{C}} \displaystyle\max_{\mathcal{D}} \mathcal{L}(\mathcal{C},\mathcal{D},t) = \mathbb{E}_{x\sim p(x)}[\text{log}(\mathcal{C}(x,t))] + \beta\mathbb{E}_{x,y\sim p(x,y)}[\text{log}(1-\mathcal{D}(\mathcal{C}(x,t),y))].
\label{eq:loss}
\end{equation}%

%% file: eqs/beta.tex
\begin{equation}
\beta(\mathcal{C},\mathcal{D},t,r) = 
\begin{cases}
    \frac{\mathbb{E}[\text{log}(\mathcal{C}(x,t-1))]_v}{\mathbb{E}[\text{log}(1-\mathcal{D}(\mathcal{C}(x,t-1),y))]_v}\cdot r & \text{if } t>0\\
    1 & \text{otherwise}
\end{cases}.
\label{eq:beta}
\end{equation}

%% file: eqs/aop.tex
\begin{equation}
\mathrm{AOP}(\lambda) = \frac{\mathrm{ACC}}{(2{\max(\mathrm{AUC}_{\text{MIA}},0.5)})^\lambda}.
\label{eq:aop}
\end{equation}

%% file: src/experiments.tex
\section{Experimental Setup}\label{Experiments}
In order to guarantee the transparency and reproducibility of the study, this section provides a comprehensive description of the experiments conducted and their setup. 
Our proposed algorithm, DAP, operates in two different settings. 
In the first one, referred to as test DAP or \textbf{DAP$_t$}, shadow models are trained with the test set, simulating a situation where an external dataset, potentially public, is accessible to both the attacker and victim. DAP$_t$ allows deep learning engineers to proactively prevent potential attacks by employing the same dataset that could be used by the attacker.
In the second setting, validation DAP or \textbf{DAP$_v$}, the shadow models are trained using the validation set. This situation mimics a typical scenario where the attacker's data distribution differs from that of the victim. 
In both of these modes, we maintained 10 shadow models, and optimize the parameter $r$ over a uniform range from 0 to 1 with increments of 0.025.

To ensure a fair evaluation, we compared DAP against several alternative approaches. We initially establish a \textbf{Baseline} model, constituting of the base classifier without any protective measures. Subsequently, following the methodology of Lomurno \etal~\cite{lomurno2023utility}, we include a model, called \textbf{Reg}, that applies dropout regularization to each intermediate classifier weight and $l2$ regularization to the model output. The dropout probability is tuned between 0.2, 0.33, and 0.5, while the $l2$ weight over 0.1, 0.01, and 0.001.
Furthermore, we extend our examination to incorporate models with ($\varepsilon$,$\delta$)-DP, retaining a constant $\delta$ value equal to 10$^{-5}$ while adjusting the $\varepsilon$ budget by modifying the number of training epochs. Specifically, we test four models with $\varepsilon$ values of 0.5, 1, 2, and 4.

\begin{figure}[t]
    \centering
    \includegraphics[width=\textwidth]{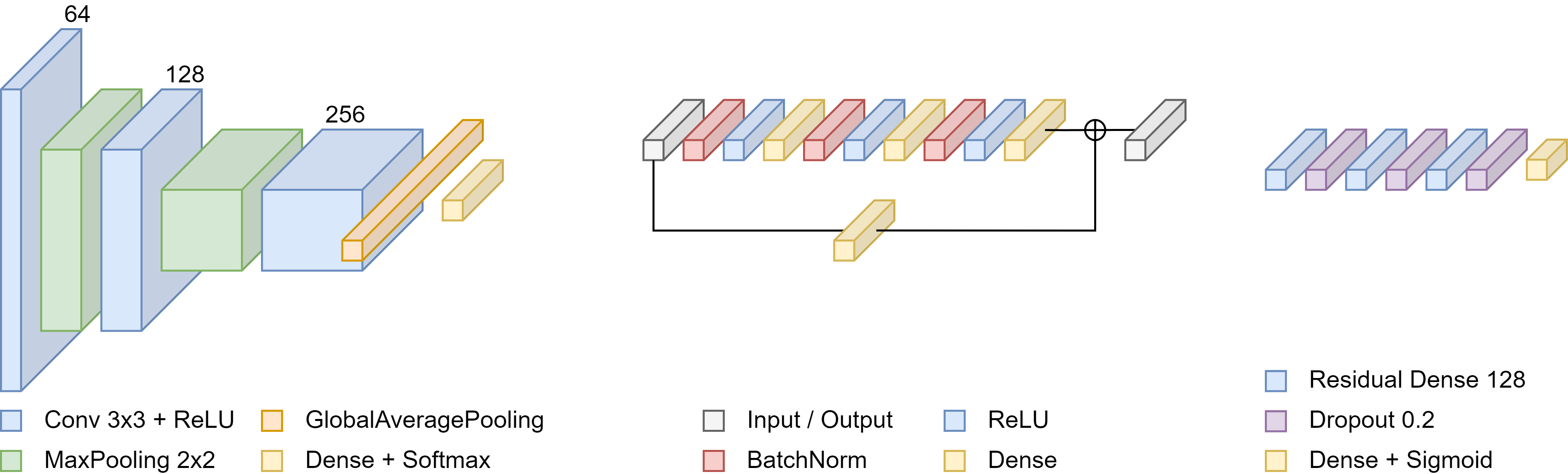}
    \caption{The neural network architectures involved in the experiments. From left to right, the CNN used for each analyzed classifier and shadow models, the residual block specially designed for the DAP discriminator, and the overall architecture of the DAP discriminator.}
    \label{fig:modules}
\end{figure}

To limit the free parameters of the experiments, we maintained the same architecture for each classifier across all configurations, as illustrated on the left side of Figure~\ref{fig:modules}. This selection was motivated by the intricate spatial complexity involved in DP training. The residual architecture of the discriminator employed in both the DAP$_t$ and DAP$_v$ models is depicted in Figure~\ref{fig:modules}, where the proposed residual block is situated in the middle, and the overall structure is positioned on the right.
All models are trained using the Adam optimizer with a learning rate chosen between $10^{-5}$, $10^{-4}$, and $10^{-3}$ and a batch size of 32. Each model is trained to convergence with early stopping -- with a patience of 25 epochs -- on validation accuracy
except for DP models, where the number of training epochs is fixed and proportional to $\varepsilon$.

The proposed models are evaluated with respect to classification Top-1 Accuracy, AUC of MIAs -- performed using the toolkit provided by the TensorFlow Privacy library -- and training epoch time. Furthermore, we employed our novel metric, the AOP, with $\lambda=2$, to assess the trade-off between performance and privacy, with a particular emphasis on the latter.
The study covers eight datasets: Cifar-10~\cite{krizhevsky2009learning}, Cifar-100~\cite{krizhevsky2009learning}, FMNIST~\cite{xiao2017fashion}, EuroSAT~\cite{helber2019eurosat}, TinyImagenet~\cite{le2015tiny}, OxfordFlowers~\cite{nilsback2008automated}, STL-10~\cite{coates2011analysis}, and Cinic-10~\cite{darlow2018cinic}. The experiments are conducted on a machine equipped with an Intel(R) Xeon(R) Gold 6238R CPU @ 2.20GHz CPU and an Nvidia Quadro RTX 6000 GPU.

%% file: src/results.tex
\section{Results and Discussion}\label{Results}

In this section, we comment on the results obtained from the set of experiments comparing DAP to regularization and DP for defense against MIAs. Table~\ref{table:acc} collects the accuracy metrics for each model and dataset. As anticipated, the models incorporating DP yield the lowest accuracy scores, even with a high privacy budget ($\varepsilon = 4$). Conversely, the regularized (Reg) model achieves consistently high accuracy, even occasionally outperforming the baseline model. Our proposed method, DAP, guaranteed an accuracy boost over the DP counterparts. A notable example of this is with the EuroSAT dataset, where the DAP$_t$ and DAP$_v$ settings result in accuracy gains of 22\% and 21\% respectively, compared to the best-performing DP model. Concerning classification accuracy, in summary, the Reg model attains the highest accuracy performance on average, followed by DAP$_t$ and DAP$_v$.

\input{tabs/accuracy}
\input{tabs/aucall}

Table~\ref{table:miss} collects the results concerning MIAs conducted on the target models. 
Here, DAP$_t$ and DAP$_v$ exhibit average AUCs of 51.3\% and 50.8\%, respectively. This indicates that both approaches effectively safeguard against MIAs, rendering the attacks nearly akin to random guessing and achieving performances competitive with DP models.
The Reg model, instead, nearly matches the privacy level of the baseline. This discrepancy between our results and the findings of Lomurno and Matteucci~\cite{lomurno2023utility} is due to the different experimental conditions. In particular, our experiments run until convergence without fixing a specific number of epochs. In summary, DAP proves to be an effective training framework that produces models resilient against MIAs.

Table~\ref{table:aop} collects the outcomes concerning the proposed AOP metric. These findings highlight the ability of DAP to produce private models that, at the same time, demonstrate competitive performance. In fact, DAP$_t$ and DAP$_v$ outperform DP and regularization in terms of the accuracy-privacy tradeoff. 
Concerning the Reg model, despite its susceptibility to MIAs, it is still a viable intermediate choice due to its superior accuracy. Conversely, DP models are extremely effective in MIA prevention but this advantage comes at the expense of the final accuracy, resulting in underperforming models. Notably, the AOP follows the trend of the privacy budget $\varepsilon$. In fact, the most private model (DP with $\varepsilon=0.5$) is also the least performing due to the impactful addition of gradient noise.

\input{tabs/aop}
\input{tabs/epochtime}

Lastly, Table~\ref{table:epochtime} collects the time per epoch required to train each model. Here, the Baseline and Reg models emerge as the fastest, while the DP models require about 8 times as long. DAP, on the other hand, manages to produce strong results both in terms of privacy and accuracy, requiring only twice the time of the baseline model.

Summarizing the results in terms of accuracy, privacy, AOP, and training time, DAP offers a better tradeoff than DP and regularization. Specifically, the DAP$_t$ setting produces high-performance models, albeit less private. In contrast, the DAP$_v$ setting produces models with strong privacy at a slight accuracy expense. Both settings handle the performance-privacy tradeoff far more effectively than DP in significantly less time.

%% file: tabs/accuracy.tex
\begin{table}[t!]
    \caption{The Accuracy metric on the test sets. Results improving the baseline are coloured in \textcolor{tabG}{green}, while results worse than the baseline are \textcolor{tabR}{red}. The best results among them are in \textbf{bold}, while the second best are \underline{underlined}.}
    \centering 
    \begin{tabularx}{\textwidth}{| l || c | Y c Y Y Y Y Y |}
    \hline
      \textbf{Dataset} & \textbf{Baseline} & \textbf{Reg}  & \textbf{${\epsilon=0.5}$} & \textbf{${\epsilon=1}$}  & \textbf{${\epsilon=2}$}  & \textbf{${\epsilon=4}$}  & \textbf{DAP$_t$}  & \textbf{DAP$_v$}\\
    \hline \hline
    \textbf{Cifar-10} & 0.784 & \textbf{\textcolor{tabG}{0.811}} & \textcolor{tabR}{0.313} & \textcolor{tabR}{0.374} & \textcolor{tabR}{0.417} & \textcolor{tabR}{0.418} & \textcolor{tabR}{\underline{0.624}} & \textcolor{tabR}{0.613}\\
    \textbf{Cifar-100} & 0.481 & \textbf{\textcolor{tabG}{0.532}} & \textcolor{tabR}{0.039} & \textcolor{tabR}{0.083} & \textcolor{tabR}{0.090} & \textcolor{tabR}{0.072} & \textcolor{tabR}{\underline{0.315}} & \textcolor{tabR}{0.276}\\
    \textbf{FMNIST} & 0.932 & \textcolor{tabR}{\textbf{0.926}} & \textcolor{tabR}{0.605} & \textcolor{tabR}{0.701} & \textcolor{tabR}{0.736} & \textcolor{tabR}{0.774} & \textcolor{tabR}{0.866} & \textcolor{tabR}{\underline{0.871}} \\
    \textbf{EuroSAT} & 0.958 & \textcolor{tabR}{\textbf{0.950}} & \textcolor{tabR}{0.308} & \textcolor{tabR}{0.588} & \textcolor{tabR}{0.681} & \textcolor{tabR}{0.646} & \textcolor{tabR}{\underline{0.900}} & \textcolor{tabR}{0.893}\\
    \textbf{TinyImagenet} & 0.365 & \textbf{\textcolor{tabG}{0.378}} & \textcolor{tabR}{0.031} & \textcolor{tabR}{0.032} & \textcolor{tabR}{0.032} & \textcolor{tabR}{0.025} & \textcolor{tabR}{\underline{0.260}} & \textcolor{tabR}{0.217}\\
    \textbf{OxfordFlowers} & 0.566 & \textbf{\textcolor{tabG}{0.659}} & \textcolor{tabR}{0.031} & \textcolor{tabR}{0.051} & \textcolor{tabR}{0.087} &\textcolor{tabR}{0.139} & \textcolor{tabR}{\underline{0.290}} & \textcolor{tabR}{0.257}\\
    \textbf{STL-10} & 0.655 & \textcolor{tabR}{\textbf{0.650}} & \textcolor{tabR}{0.084} & \textcolor{tabR}{0.142} & \textcolor{tabR}{0.250} & \textcolor{tabR}{0.289} & \textcolor{tabR}{\underline{0.480}} & \textcolor{tabR}{0.384}\\
    \textbf{Cinic-10} & 0.673 & \textbf{\textcolor{tabG}{0.709}} & \textcolor{tabR}{0.280} & \textcolor{tabR}{0.341} & \textcolor{tabR}{0.391} & \textcolor{tabR}{0.405} & \textcolor{tabR}{0.577} & \textcolor{tabR}{\underline{0.586}}\\
    \hline
    \textbf{Average} & 0.677 & \textbf{\textcolor{tabG}{0.702}} & \textcolor{tabR}{0.211} & \textcolor{tabR}{0.289} & \textcolor{tabR}{0.336} & \textcolor{tabR}{0.346} & \textcolor{tabR}{\underline{0.539}} & \textcolor{tabR}{0.512}\\
    \hline
    \end{tabularx}
    \label{table:acc}
\end{table}

%% file: tabs/aucall.tex
\begin{table}[t!]
    \caption{The AUC metric of the MIAs. Results improving the baseline are coloured in \textcolor{tabG}{green}, while results worse than the baseline are \textcolor{tabR}{red}. The best results among them are in \textbf{bold}, while the second best are \underline{underlined}.}
    \centering 
    \resizebox{\textwidth}{!}{%
    \begin{tabularx}{\textwidth}{| l || c | Y c Y Y Y Y Y |}
    \hline
    \textbf{Dataset} & \textbf{Baseline} & \textbf{Reg}  & \textbf{${\epsilon=0.5}$} & \textbf{${\epsilon=1}$}  & \textbf{${\epsilon=2}$}  & \textbf{${\epsilon=4}$}  & \textbf{DAP$_t$}  & \textbf{DAP$_v$}\\
    \hline \hline
    \textbf{Cifar-10} & 0.648 & \textcolor{tabG}{0.631} & \textcolor{tabG}{\underline{0.505}} & \textcolor{tabG}{0.526} & \textcolor{tabG}{0.519} & \textbf{\textcolor{tabG}{0.503}} & \textcolor{tabG}{0.507} & \underline{\textcolor{tabG}{0.505}}\\
    \textbf{Cifar-100} & 0.603 & \textcolor{tabR}{0.621} & \textbf{\textcolor{tabG}{0.501}} & \textcolor{tabG}{0.515} & \textcolor{tabG}{0.507} & \textcolor{tabG}{\underline{0.506}} & \textcolor{tabG}{0.516} & \textcolor{tabG}{\underline{0.506}}\\
    \textbf{FMNIST} & 0.552 & \textcolor{tabR}{0.562} & \textbf{\textcolor{tabG}{0.502}} & \textbf{\textcolor{tabG}{0.502}} & \textcolor{tabG}{\underline{0.504}} & \textcolor{tabG}{0.505} & \textcolor{tabG}{0.507} & \textcolor{tabG}{0.506} \\
    \textbf{EuroSAT} & 0.544 & \textcolor{tabG}{0.528} & \textcolor{tabG}{0.505} & \textcolor{tabG}{0.502} & \textbf{\textcolor{tabG}{0.500}} & \textcolor{tabG}{0.502} & \textcolor{tabG}{\underline{0.501}} & \textcolor{tabG}{\underline{0.501}}\\
    \textbf{TinyImagenet} & 0.603 & \textcolor{tabG}{0.592} & \textcolor{tabG}{0.514} & \textbf{\textcolor{tabG}{0.501}} & \textcolor{tabG}{0.521} & \textcolor{tabG}{\underline{0.504}} & \textcolor{tabG}{0.516} & \textcolor{tabG}{0.509}\\
    \textbf{OxfordFlowers} & 0.761 & \textcolor{tabR}{0.765} & \textcolor{tabG}{0.543} & \textcolor{tabG}{0.537} & \textcolor{tabG}{\underline{0.526}} & \textcolor{tabG}{0.532} & \textcolor{tabG}{0.538} & \textcolor{tabG}{\textbf{0.521}}\\
    \textbf{STL-10} & 0.604 & \textcolor{tabG}{0.563} & \textcolor{tabG}{\underline{0.502}} & \textcolor{tabG}{0.524} & \textcolor{tabG}{0.505} & \textbf{\textcolor{tabG}{0.501}} & \textcolor{tabG}{0.508} & \textcolor{tabG}{0.506}\\
    \textbf{Cinic-10} & 0.572 & \textcolor{tabR}{0.614} & \textbf{\textcolor{tabG}{0.501}} & \textcolor{tabG}{0.514} & \textcolor{tabG}{0.511} & \underline{\textcolor{tabG}{0.507}} & \textcolor{tabG}{0.513} & \textcolor{tabG}{0.507}\\
    \hline
    \textbf{Average} & 0.611 & \textcolor{tabG}{0.609} & \textcolor{tabG}{0.509} & \textcolor{tabG}{0.514} & \textcolor{tabG}{0.511} & \textbf{\textcolor{tabG}{0.507}} & \textcolor{tabG}{0.513} & \textcolor{tabG}{\underline{0.508}}\\
    \hline
    \end{tabularx}}%
    \label{table:miss}
\end{table}

%% file: tabs/aop.tex
\begin{table}[t!]
    \caption{The AOP metric on the test sets. Results improving the baseline are coloured in \textcolor{tabG}{green}, while results worse than the baseline are \textcolor{tabR}{red}. The best results among them are in \textbf{bold}, while the second best are \underline{underlined}.}
    \centering 
    \resizebox{\textwidth}{!}{%
    \begin{tabularx}{\textwidth}{| l || c | Y c Y Y Y Y Y |}
    \hline
      \textbf{Dataset} & \textbf{Baseline} & \textbf{Reg}  & \textbf{${\epsilon=0.5}$} & \textbf{${\epsilon=1}$}  & \textbf{${\epsilon=2}$}  & \textbf{${\epsilon=4}$}  & \textbf{DAP$_t$}  & \textbf{DAP$_v$}\\
    \hline \hline
    \textbf{Cifar-10} & 0.467 &	\textcolor{tabG}{0.509} &	\textcolor{tabR}{0.307} &	\textcolor{tabR}{0.338} & \textcolor{tabR}{0.387} & \textcolor{tabR}{0.413}	& \textbf{\textcolor{tabG}{0.607}} &	\textcolor{tabG}{\underline{0.601}}\\
    \textbf{Cifar-100} & 0.331	&\textbf{\textcolor{tabG}{0.345}}	&\textcolor{tabR}{0.039}	&\textcolor{tabR}{0.078}	&\textcolor{tabR}{0.087}	&\textcolor{tabR}{0.070}&	\textcolor{tabR}{\underline{0.296}}	& \textcolor{tabR}{0.269}\\
    \textbf{FMNIST} & 0.765	& \textcolor{tabR}{0.733} & \textcolor{tabR}{0.600}& \textcolor{tabR}{0.695}& \textcolor{tabR}{0.724} &	\textcolor{tabR}{0.759} &	\textcolor{tabG}{\underline{0.842}} & \textcolor{tabG}{\textbf{0.850}} \\
    \textbf{EuroSAT} & 0.809	& \textcolor{tabG}{0.852}	& \textcolor{tabR}{0.302}	& \textcolor{tabR}{0.583} &\textcolor{tabR}{0.681} & \textcolor{tabR}{0.641} 	& \textbf{\textcolor{tabG}{0.896}} &	\textcolor{tabG}{\underline{0.889}} \\
    \textbf{TinyImagenet} & 0.251 &	\textbf{\textcolor{tabG}{0.270}} &	\textcolor{tabR}{0.029} &	\textcolor{tabR}{0.032} &	\textcolor{tabR}{0.029} &	\textcolor{tabR}{0.025} &	\textcolor{tabR}{\underline{0.244}}& \textcolor{tabR}{0.209}\\
    \textbf{OxfordFlowers} & 0.244 & \textcolor{tabG}{\textbf{0.281}} &	\textcolor{tabR}{0.026} &	\textcolor{tabR}{0.044} &	\textcolor{tabR}{0.079} &	\textcolor{tabR}{0.123} &	\textcolor{tabG}{\underline{0.250}} &	\textcolor{tabR}{0.237} \\
    \textbf{STL-10} & 0.449 &	\textbf{\textcolor{tabG}{0.513}} &	\textcolor{tabR}{0.083} &	\textcolor{tabR}{0.129} & \textcolor{tabR}{0.245} & \textcolor{tabR}{0.288} &	\textcolor{tabG}{\underline{0.465}}& \textcolor{tabR}{0.375} \\
    \textbf{Cinic-10} & 0.514 & \textcolor{tabR}{0.470} & \textcolor{tabR}{0.279} &	\textcolor{tabR}{0.337} & \textcolor{tabR}{0.386} &	\textcolor{tabR}{0.399} &	\textcolor{tabG}{\underline{0.552}}	& \textcolor{tabG}{\textbf{0.570}}\\
    \hline
    \textbf{Average} & 0.479 & \textcolor{tabG}{0.497} & \textcolor{tabR}{0.208} &	\textcolor{tabR}{0.280} &	\textcolor{tabR}{0.327} & \textcolor{tabR}{0.340}  & \textbf{\textcolor{tabG}{0.519}} & \textcolor{tabG}{\underline{0.500}}\\
    \hline
    \end{tabularx}}%
    \label{table:aop}
\end{table}

%% file: tabs/epochtime.tex
\begin{table}[t!]
    \caption{Training time per epoch required for each experiment, measured in seconds. Results improving the baseline are coloured in \textcolor{tabG}{green}, while results worse than the baseline are \textcolor{tabR}{red}. The best results among them are in \textbf{bold}, while the second best are \underline{underlined}.}
    \centering 
    \begin{tabularx}{\textwidth}{| l || c | Y Y Y Y |}
    \hline
      \textbf{Dataset} & \textbf{Baseline} & \textbf{Reg}  & \textbf{DP} & \textbf{DAP$_t$}  & \textbf{DAP$_v$}\\
    \hline \hline
    \textbf{Cifar-10} & 5.6 & \textcolor{tabR}{\textbf{5.9}} & \textcolor{tabR}{46.5} & \textcolor{tabR}{17.8} & \textcolor{tabR}{\underline{17.7}}\\
    \textbf{Cifar-100} & 8.9 & \textcolor{tabR}{\textbf{9.5}} & \textcolor{tabR}{48.2} & \textcolor{tabR}{\underline{17.9}} & \textcolor{tabR}{\underline{17.9}}\\
    \textbf{FMNIST} & 12.1 & \textcolor{tabG}{\textbf{11.7}} & \textcolor{tabR}{53.2} & \textcolor{tabR}{\underline{23.4}} & \textcolor{tabR}{23.5}\\
    \textbf{EuroSAT} & 4.7 & \textcolor{tabR}{\textbf{5.5}} & \textcolor{tabR}{72.7} & \textcolor{tabR}{10.2} & \textcolor{tabR}{\underline{9.5}}\\
    \textbf{TinyImagenet} & 31.7 & \textcolor{tabR}{\textbf{32.6}} & \textcolor{tabR}{338.5} & \textcolor{tabR}{62.4} & \textcolor{tabR}{\underline{61.4}}\\
    \textbf{OxfordFlowers} & 1.2 & \textcolor{tabR}{\textbf{1.8}} & \textcolor{tabR}{20.8} & \textcolor{tabR}{\underline{2.3}} & \textcolor{tabR}{2.4}\\
    \textbf{STL-10} & 1.7 & \textcolor{tabR}{\textbf{2.5}} & \textcolor{tabR}{34.9} & \textcolor{tabR}{\underline{2.5}} & \textcolor{tabR}{2.8}\\
    \textbf{Cinic-10} & 30.0 & \textcolor{tabG}{\textbf{22.1}} & \textcolor{tabR}{106.5} & \textcolor{tabR}{\underline{41.2}} & \textcolor{tabR}{42.8}\\
    \hline
    \textbf{Average} & 12.0 & \textcolor{tabG}{\textbf{11.5}} & \textcolor{tabR}{90.2} & \textcolor{tabR}{\underline{22.2}} & \textcolor{tabR}{\underline{22.2}}\\
    \hline
    \end{tabularx}
    \label{table:epochtime}
\end{table}

%% file: src/conclusion.tex
\section{Conclusion}\label{Conclusion}
In this work, we introduced the Discriminative Adversarial Privacy (DAP) framework and the Accuracy Over Privacy (AOP) metric. The goal of DAP is to produce deep learning models resilient to Membership Inference Attacks (MIAs), while the AOP summarizes the privacy-accuracy tradeoff in a single value. As shown in the experiments, DAP demonstrated superior ability in maintaining a beneficial balance between model performance and privacy, outperforming models based on Differential Privacy (DP). The AOP metric has effectively encapsulated these results, providing a concise yet robust evaluation criterion. On top of that, DAP required considerably less computational overhead, thus accelerating the training process with respect to DP. Collectively, our contributions offer a promising approach to the development and evaluation of deep learning models resilient against MIAs, providing an optimal balance between execution time, accuracy, and privacy.

%% file: src/acknowledgements.tex
\section*{Acknowledgment}\label{sec:ack}
This project has been supported by AI-SPRINT: AI in Secure Privacy-pReserving computINg conTinuum (European Union H2020 grant agreement No. 101016577) and FAIR: Future Artificial Intelligence Research (NextGenerationEU, PNRR-PE-AI scheme, M4C2, investment 1.3, line on Artificial Intelligence).